
\magnification=1200
\hsize=6truein\vsize=8.5truein
\font\open=msbm10 

\font\bigbf=cmbx10 scaled\magstep1

\def\mbox#1{{\leavevmode\hbox{#1}}}
\def\hspace#1{{\phantom{\mbox#1}}}
\def\oR{\mbox{\open\char82}}
\def\oZ{\mbox{\open\char90}}

\def\rS{{\rm S}}

\def\la{\lambda}
\def\si{\sigma}
\def\det{{\rm det\,}}
\def\Real{{\rm Re\,}}
\def\zf{$\zeta$--function}
\def\zfs{$\zeta$--functions}

\def\frac#1/#2{\leavevmode\kern.1em
\raise.5ex\hbox{\the\scriptfont0 #1}\kern-.1em/\kern-.15em
\lower.25ex\hbox{\the\scriptfont0 #2}}
\def\sfrac#1/#2{\leavevmode\kern.1em
\raise.5ex\hbox{\the\scriptscriptfont0 #1}\kern-.1em/\kern-.15em
\lower.25ex\hbox{\the\scriptscriptfont0 #2}}

\def\gtorder{\mathrel{\raise.3ex\hbox{$>$}\mkern-14mu
             \lower0.6ex\hbox{$\sim$}}}
\def\ltorder{\mathrel{\raise.3ex\hbox{$<$}|mkern-14mu
             \lower0.6ex\hbox{\sim$}}}

\def\semidirprod{\rlap{\ss C}\raise1pt\hbox{$\mkern.75mu\times$}}

\def\for{\lower6pt\hbox{$\Big|$}}
\def\fish{\kern-.25em{\phantom{abcde}\over \phantom{abcde}}\kern-.25em}

\def\boxit#1{\vbox{\hrule\hbox{\vrule\kern3pt
        \vbox{\kern3pt#1\kern3pt}\kern3pt\vrule}\hrule}}
\def\dalemb#1#2{{\vbox{\hrule height .#2pt
        \hbox{\vrule width.#2pt height#1pt \kern#1pt
                \vrule width.#2pt}
        \hrule height.#2pt}}}


\def\noin{\noindent}

\def\al{\alpha}
\def\be{\beta}
\def\ga{\gamma}

\def\Ga{\Gamma}

\def\ka{\kappa}
\def\la{\lambda}

\def\si{\sigma}

\def\th{\theta}
\def\ze{\zeta}

\def\comb#1#2{{\left(#1\atop#2\right)}}

\def\eg{{\it e.g. }}
\def\ie{{\it i.e. }}
\def\cf{{\it cf }}


\def\gap{\vskip 20truept}

\def\sect{{\vskip 10truept\noindent}}

\def\3j#1#2#3#4#5#6{\left\lgroup\matrix{#1&#2&#3\cr#4&#5&#6\cr}
\right\rgroup}


\def\nolabels{\def\eqnlabel##1{}\def\eqlabel##1{}\def\reflabel##1{}}
\def\writelabels{\def\eqnlabel##1{%
{\escapechar=` \hfill\rlap{\hskip.09in\string##1}}}%
\def\eqlabel##1{{\escapechar=` \rlap{\hskip.09in\string##1}}}%
\def\reflabel##1{\noexpand\llap{\string\string\string##1\hskip.31in}}}
\nolabels
\global\newcount\meqno \global\meqno=1
\global\meqno=1
\def\eqnn#1{\xdef #1{(\the\meqno)}%
\global\advance\meqno by1\eqnlabel#1}
\def\eqna#1{\xdef #1##1{\hbox{$(\the\meqno##1)$}}%
\global\advance\meqno by1\eqnlabel{#1$\{\}$}}
\def\eqn#1#2{\xdef #1{(\the\meqno)}\global\advance\meqno by1%
$$#2\eqno#1\eqlabel#1$$}


\global\newcount\refno
\global\refno=1 \newwrite\reffile
\newwrite\refmac
\newlinechar=`\^^J
\def\ref#1#2{\the\refno\nref#1{#2}}
\def\nref#1#2{\xdef#1{{\bf\the\refno}} 
\ifnum\refno=1\immediate\openout\reffile=refs.tmp\fi
\immediate\write\reffile{
     \noexpand\item{[{\noexpand#1}]\ }#2\noexpand\nobreak.}
     \immediate\write\refmac{\def\noexpand#1{\the\refno}}
   \global\advance\refno by1}
\def\semi{;\hfil\noexpand\break ^^J}
\def\refn#1#2{\nref#1{#2}}

\def\anp#1#2#3{{\it Ann. Phys.} {\bf {#1}} (19{#2}) #3}

\def\cmp#1#2#3{{\it Comm. Math. Phys.} {\bf {#1}} (19{#2}) #3}

\def\jmp#1#2#3{{\it J. Math. Phys.} {\bf {#1}} (19{#2}) #3}

\def\np#1#2#3{{\it Nucl. Phys.} {\bf B{#1}} (19{#2}) #3}
\def\pl#1#2#3{{\it Phys. Lett.} {\bf {#1}B} (19{#2}) #3}

\def\pr#1#2#3{{\it Phys. Rev.} {\bf {#1}} (19{#2}) #3}

\def\prD#1#2#3{{\it Phys. Rev.} {\bf D{#1}} (19{#2}) #3}

\def\mz#1#2#3{{\it Math. Zeit.} {\bf {#1}} ({#2}) #3}
\def\pams#1#2#3{{\it Proc. Am. Math. Soc.} {\bf {#1}} (19{#2}) #3}

\def\top#1#2#3{{\it Topology} {\bf {#1}} (19{#2}) #3}
\def\tams#1#2#3{{\it Trans. Am. Math. Soc.} {\bf {#1}} (19{#2}) #3}

\refn\Dowka{J.S.Dowker {\it Effective action in spherical domains}. {\it Comm.
Math. Phys.} to be published}
\refn\Aurell{E.Aurell and P.Salomonson {\it On functional determinants of
Laplacians in polygons and simplices}}
\refn\Branson{T.P.Branson and B.\O rsted \pams{113}{91}{669}}
\refn\Quine{J.R.Quine, S.H.Heydari and R.Y.Song \tams{338}{93}{213}}
\refn\Minak{S.Minakshisudaram {\it J. Ind. Math. Soc.} {\bf 13} (1949) 41}
\refn\Camporesi{R.Camporesi {\it Physics Reports} {\bf 196} (1990) 1}
\refn\Cand{P.Candelas and S.Weinberg \np{237}{84}{397}}
\refn\Chodos{A.Chodos and E.Myers \anp{156}{84}{412 }}
\refn\Barnesa{E.W.Barnes {\it Trans. Camb. Phil. Soc.} {\bf 19} (1903) 374}
\refn\Dowkb{J.S.Dowker \pr{36}{86}{1111}}
\refn\Barnesb{E.W.Barnes {\it Trans. Camb. Phil. Soc.} {\bf 19} (1903) 426}
\refn\Hortacsu{M.Hortacsu, K.D.Rothe and B.Schroer \prD {20}{79}{3203}}
\refn\Weisbergera {W.I.Weisberger \np{B284}{87}{171}}
\refn\Voros{A.Voros \cmp{110}{87}{110}}
\refn\Vardi{I.Vardi {\it SIAM J. Math. Anal.} {\bf 19} (1988) 493}
\refn\Gillet{H.Gillet, C.Soul\'e and D.Zagier \top{30}{92}{21}}
\refn\Nash{C.Nash and D.J.O'Connor {\it Determinants of Laplacians, the
Ray-Singer torsion on lens spaces and the Riemann zeta function}}
\refn\Weisbergerb {W.I.Weisberger \cmp{112}{87}{633}}
\refn\Dowkc{J.S.Dowker {\it Functional determinants on regions of the
plane and sphere}. {\it Class. Quant. Grav.} to be published}
\refn\DandS{J.S.Dowker and J.P.Schofield \jmp{31}{90}{808}}
\refn\Gromes{D.Gromes \mz{94}{1966}110}
\refn\Pockels{F.Pockels {\it \"Uber die partielle differentialgleichung
$\Delta u+k^2u=0$}, B.G.Teubner, Leipzig 1891}
\refn\Luscher{M.L\"uscher, K.Symanzik and P.Weiss \np {173}{80}{365}}
\refn\Polyakov{A.M.Polyakov \pl {103}{81}{207}}
\refn\Bukhb{L.Bukhbinder, V.P.Gusynin and P.I.Fomin {\it Sov. J. Nucl.
 Phys.} {\bf 44} (1986) 534}
\refn\Alvarez{O.Alvarez \np {216}{83}{125}}
%


\vglue 1truein
\rightline {MUTP/93/28}
\gap
\centerline {\bigbf Functional determinants on spheres and sectors}
\vskip 15truept
\centerline{J.S.Dowker}
\vskip 10 truept
\centerline {\it Department of Theoretical Physics,}
\centerline{\it The University of Manchester, Manchester, England.}
\vskip 40truept
\centerline {Abstract}
\vskip 10truept
\noin The scalar functional determinants on sectors of the two-dimensional disc
and spherical cap are
determined for arbitrary angles (rational factors of $\pi$). The wholesphere
and hemisphere expressions are also given, in low dimensions, for both the
ordinary and the conformal Laplacian.

\rightline{December 1993}
\vskip 10truept
\vfill\eject
\sect{\bf 1. Introduction.}

This paper is a direct continuation of a previous one, [\Dowka], where we
discussed the effective, one-loop action on those regions of the sphere that
are fundamental domains of the finite subgroups of O(3). The simplest of these
are the dihedral and double dihedral groups, corresponding to the lune and
halflune respectively. The angle, $\be$, of the lune equals $\pi/q$,
with $q$ integral.

Of course the eigenvalues of the Laplacian on a lune can be found
for any angle and it is this case we wish to look at here. A
stereographic projection enables the functional determinant,
$\det(-\nabla^2)$,
on a disc sector of angle $\be$ to be found, as well as that on
the sector of a spherical cap -- an equilateral spherical triangle,
$(2,2,\pi/\be)$.

An interesting open problem is the determination of $\det(-\nabla^2)$
on a general geodesic spherical triangle, to parallel the recent result of
Aurell and Salomonson [\Aurell] for planar triangles.

The evaluation of functional determinants is something of an aesthetic
computational puzzle, involving eigenvalue finding and analytical
continuation. Only certain manifolds can be treated efficiently or
prettily and it is in this context that the following results
are presented.

The most interesting systems are the hyperbolic ones, with the positive
curvature case being trivial, in some sense. For this reason, there exists
less work on spherical domains. However, in our opinion, there are still
some attractive areas worth exploring.

We begin with a discussion of functional determinants on the {\it complete}
sphere. Although our concern is really with regions of the sphere and plane,
a re-examination of the wholesphere is relevant in view of recent work by
mathematicians, [\Branson, \Quine]. In [\Dowka] we rederived known results on
the whole two-sphere as a byproduct of the formulae on the quotient
$\rS^2/\Ga$, in particular on the hemisphere.
Our intention here is to continue this to the higher spheres more explicitly.
Attention is restricted to minimal coupling, \ie to the \zf\ of the
Laplacian, and to conformal coupling, \ie to
$\det\big(-\nabla^2+R(d-2)/4(d-1)\big)$.
\sect{\bf 2. The wholesphere and the hemisphere.}

\noin The \zf\ on spheres has been studied for a long time [\Minak] and has
benefitted from a detailed examination over the past fifteen
years or so, \eg [\Camporesi]. Quantities calculated are, for example, the
conformal
anomaly and the one-loop vacuum energy. These involve the evaluation of the
\zf\ at negative half-integers. Our interest here is with the functional
determinant.

Define the \zf, \cf [\Dowka],
\eqn\genzetaa{
\ze(s,a,\al;{\bf r})=\sum_{\bf m=0}^\infty {1\over\big((a+{\bf m.r})^2-
\al^2\big)^s}
}
where ${\bf r}$ and ${\bf m}$ are
$d$-vectors and $\al$ equals $(d-1)/2$ for minimal coupling and $1/2$ for
conformal. The \zf\ on a portion of the $d$-sphere,
$\rS^d/\Ga$, corresponds to the values $a=(d+1)/2$ for Dirichlet
conditions,  and $a=(d-1)/2$ for Neumann conditions. For the latter, the
term ${\bf m}=(0,0,\ldots,0)$ is to be omitted from the sum for minimal
coupling. If ${\bf r}=(1,1,\ldots,1)$, giving the hemisphere, the sum
over the ${\bf m}$ can be done to yield the general form,
\eqn\hemizeta{
\ze(s)=\sum_{m=0}^\infty\comb{m+d-1}{d-1}{1\over\big((a+m)^2-
\al^2)\big)^s}.
}

We may show that, as can be anticipated, the whole \zf\ is the
sum of the hemisphere Dirichlet and Neumann \zfs. Translating the
Dirichlet summation integer by unity, we get, concentrating on minimal
coupling,
\eqn\sumzeta{\eqalign{
\ze_N(s)+\ze_D(s)&=\sum_{m=1}^\infty\left[\comb{m+d-1}{d-1}+\comb{m+d-2}
{d-1}\right]{1\over\big(m(m+d-1)\big)^s}\cr
&=\sum_{m=1}^\infty\left[\comb{m+d}{d}-\comb{m+d-2}
{d}\right]{1\over\big(m(m+d-1)\big)^s}\cr
&=\ze_{\rm rot}(s)\,,\cr
}
}
after some minor rearrangement. The standard eigenvalues and degeneracies
can be recognised in this. They are very old.

A traditional way, [\Minak, \Cand, \Chodos], of dealing with this
expression is to write the eigenvalues as in \hemizeta\ and use the Bessel
function identity
\eqn\bessel{
{1\over{(z^2-\al^2)}^s}={\sqrt\pi\over\Ga(s)}
\int_0^\infty e^{-zt}\left({t\over2\al}\right)^{s-1/2}I_{s-1/2}(\al t)\,dt
}
to allow the sum over $m$ to be performed. Pursuing this path involves a
mild analytic continuation and would doubtless ultimately lead to our
goal.
For present purposes we prefer the method in [\Dowka] which is
the reasonably standard one of expanding in powers of $\al$. This
produces expressions in terms of the Barnes \zf, [\Barnesa]
\eqn\Barnes{
\ze_d\big(s,a\mid{\bf r}\big)=\sum_{\bf m=0}^\infty{1\over(a+{\bf m.r})^s}
\,,\quad \Real s>d.
}
For ${\bf r=1}$,
\eqn\hemiBarnes{\eqalign{
\ze_d(s,a)&={i\Ga(1-s)\over2\pi}\int_L{e^{z(d/2-a)}(-z)^{s-1}\over
2^d\sinh^d(z/2)}\,dz\cr
&=\sum_{m=0}^\infty\comb{m+d-1}{d-1}{1\over{(a+m)}^s}\,,\quad \Real s>d.\cr
}
}
$L$ is the Hankel contour. As usual, if $a=0$, {\it it will be understood}
that the $m=0$ term is not included in the sum.

With this convention in mind,
the formulae derived in [\Dowka] allow one to write,
\eqn\neu{
\ze_N'(0)=\ze_d'(0,0)+\ze_d'(0,d-1)+\ln(d-1)-\sum_{r=1}^u{\al^{2r}\over r}
N^N_{2r}(d)\sum_{k=0}^{r-1}{1\over 2k+1}
}
\eqn\dir{
\ze_D'(0)=\ze_d'(0,1)+\ze_d'(0,d)-\sum_{r=1}^u{\al^{2r}\over r}
N^D_{2r}(d)\sum_{k=0}^{r-1}{1\over 2k+1}.
}
Here $u$ equals $d/2$ if $d$ is even, and $(d-1)/2$ if $d$ is odd.
$N$ is the residue of the Barnes \zf,
\eqn\residue{
\ze_d(s+r,a)\rightarrow{N_r(d)\over s}\quad{\rm as}\,\,s\rightarrow0.
}
$N$ depends on $a$ and is given by a generalized Bernoulli polynomial. In the
present case it is easiest to find the residue directly from the integral in
\hemiBarnes. This has earlier appeared [\Dowkb] in connection with
cosmic string calculations and should also emerge from \bessel, [\Chodos].

In accordance with our previous result, \sumzeta, the wholesphere value is
obtained by adding \neu\ and \dir.
The final terms produce the combination $N^N+N^D$ which, from
the poles in the $\Ga$-function in \hemiBarnes, equals
\eqn\residuesuma{\eqalign{
N_{2r}^N(d)+N_{2r}^D(d)&=
{i2^{2r-d+1}\over2\pi (2r-1)!}\oint_C{z^{2r-1}\cosh z\over\sinh^d\!z}\,dz\cr
&={2^{2r-d+1}\over2\pi i(d-1)(2r-2)!}\oint_C{z^{2r-2}\over\sinh^{d-1}\!z}\,
dz \cr
}
}
where $C$ is a small contour around the origin and $z$ has been rescaled by
a factor of two. If $d$ is odd, the integrand is
even and the integral is zero. If $d$ is even we pick out the term proportional
to $1/z$ using the standard expansion
\eqn\expsn{
\left({z\over\sinh z}\right)^{d-1}=\sum_{\nu=0}^\infty D^{(d-1)}_
{2\nu}{z^{2\nu}\over(2\nu)!}
}
to get the residue $N^{\rm rot}_{2r}(d)$, $=N_{2r}^N(d)+N_{2r}^D(d)$,
\eqn\residuesumb{
N^{\rm rot}_{2r}(d)=
{2^{2r-d+1}\over(d-1)(2r-2)!(d-2r)!}D^{(d-1)}_{d-2r},\quad
1\le r\le d/2.
}
in terms of the $D$-coefficients related to generalised Bernoulli functions.

An explicit realisation suitable for symbolic manipulation is
\eqn\residuesumc{
N^{\rm rot}_{2r}(d)=
{2\over(d-1)!(2r-2)!}\,{d^{2r-2}\over dx^{2r-2}}\,
\prod_{i=1}^{d-2}(x-i)\,\bigg|_{{}_{x=(d-1)/2}}.
}

We next deal with the four \zf\ derivatives. Adjustment
of the lower limit of the summations, where necessary, gives
$$
\ze_d(s,0)+\ze_d(s,d-1)+\ze_d(s,1)+\ze_d(s,d)=
$$
\eqn\fourzetas{\sum_{m=1}^\infty\left[\comb{m+d-1}{d-1}+\comb{m+d-2}{d-1}+
\comb{m-1}{d-1}+\comb{m}{d-1}\right]{1\over m^s}.
}
Any undefined factorials are to be defined by $\Ga$-functions and
cancellations have occurred.

The combination of factorials is rewritten
$$
\comb{m}{d-1}-(-1)^d\comb{-m}{d-1}+\comb{m-1}{d-1}-(-1)^d\comb{-m-1}{d-1},
$$
which is a polynomial in $m$, even for odd $d$ and odd for even, and can be
expressed in terms of Stirling numbers, $S^{(k)}_j$. Some relevant expansions
are also given in [\Barnesb]. An alternative to expanding the factorials
is to work directly from the integral in \hemiBarnes.

The derivative of
\fourzetas\ at $s=0$ can thus be written as a sum of derivatives of
Riemann \zfs\ evaluated at nonpositive integers.

Putting all the pieces together, we obtain the following formula enabling the
regularised functional determinant on the complete, unit $d$-sphere,
$e^{-\ze_{\rm rot}'(0)}$, to be found,

\eqn\zetadash{\eqalign{
\ze_{\rm rot}'(0)&=
{1\over(d-1)!}\sum_{k=0}^{d-1}\big(1-(-1)^{d+k}\big)\big(S^{(k)}_{d-1}
+S^{(k+1)}_d\big)\ze_R'(-k)\cr
&\quad-\sum_{r=1}^u{(d-1)^{2r}\over2^{2r}r}
N^{\rm rot}_{2r}(d)\sum_{k=0}^{r-1}{1\over 2k+1} +\ln(d-1)\,.\cr
}
}

Some low-dimensional expressions for $\ze_{\rm rot}'(0)$ are
\eqn\resultsa{\eqalign{
d=2,\quad\quad&4\ze_R'(-1)-{1\over2}\approx-1.1617\cr
d=3,\quad\quad&2\ze_R'(-2)+2\ze_R'(0)+\ln2\approx-1.2056\cr
d=4,\quad\quad&{2\over3}\ze_R'(-3)+{13\over3}\ze_R'(-1)-{15\over16}+\ln3
\approx-0.5521\cr
d=5,\quad\quad&{1\over6}\ze_R'(-4)+{23\over6}\ze_R'(-2)+2\ze_R'(0)+\ln4
\approx-0.5670\cr
d=6,\quad\quad&{1\over30}\ze_R'(-5)+2\ze_R'(-3)+{149\over30}\ze_R'(-1)-
{455\over432}+\ln5\approx-0.2547\cr
d=7,\quad\quad&{1\over180}\ze_R'(-6)+{13\over18}\ze_R'(-4)+{949\over180}
\ze_R'(-2)+2\ze_R'(0)+\ln6\approx-0.2009.\cr
}
}
Figure 1 gives a plot of the values and shows a different behaviour
for odd and even dimensions. Perhaps this is not unexpected for low dimensions
but one might have hoped that the values would converge for large $d$. A
curiosity is the crossing exactly on the axis around $d\approx8.5$, allowing
for the interpolation.

The two-sphere result appears to have been derived first by Horta\c{c}su et al
[\Hortacsu] and later by Weisberger [\Weisbergera]. Another treatment is given
by Voros [\Voros] as an example of a general method of evaluating functional
determinants. (There is a minor misprint in equation (6.41) of this
reference.)

The paper by Vardi [\Vardi] also contains a method of evaluating these
determinants. The basic calculational formula is Proposition 3.1, which
we repeat here. Defining
\eqn\hdef{
H_p(s)=\sum_{m=1}^\infty{m^p\over\big(m(m+n)\big)^s}
}
then
\eqn\hdash{
H'_p(0)=\sum_{k=1}^n(k-n)^p\ln k-{(-n)^{p+1}\over2(p+1)}
\sum_{j=1}^p{1\over j}+\ze'_R(-p)+
(-n)^p\sum_{r=0}^p\comb{p}{r}{\ze'_R(-r)\over(-n)^r}.
}
The case when $p=1$ was already given by Horta\c csu {\it et al}
 [\Hortacsu]. This result has been used by Gillet {\it et al} [\Gillet]
in connection with analytic torsion and the arithmetic Todd genus.

The \zf\ on the $d$-sphere is built out of the $H_p$ for various $p$. In the
case of the 3-sphere we have checked that it gives the result in \resultsa\
and not that exhibited in [\Vardi], {\it viz.}
$2\ze_R'(-2)-\ze_R'(-1)+3\ze_R'(0)$.
Other values of $d$ have also been checked. We conclude that there is some
arithmetic error in Vardi's substitutions. This renders the particular
relation between $\Ga_3(1/2)$ and the functional determinants in Theorem
1.1 of [\Vardi] incorrect. It seems the relation should read
$\Ga_3(1/2)=\det\Delta_3^{7/16}\det\Delta_2^{3/8}\det\Delta_1^{-1/8}
2^{5/12}\exp(-3/16)\approx1.831123$.

Confirmation of our value of $\ze'_{\rm rot}(0)$ can be found in the
calculations of Nash
and O'Connor [\Nash], Appendix A, who, incidental to a discussion of
analytic torsion, obtained $-1.20563$ by somewhat roundabout means.

In this connection, consider the \zf
\eqn\betazeta{
\ze(s,\be)=\sum_{m=1}^\infty{m^2\over(m^2-\be^2)^s}.
}
When $\be=1$, and the sum runs from 2 upwards, this is $\ze_{\rm rot}$ on the
three-sphere, \sumzeta. The method in [\Dowka] can be applied directly to
\betazeta\ and leads to
\eqn\difbetazeta{
\ze'(0,\be)=2\ze_R'(-2)-\be^2\ln\sin(\pi\be)+
2\int_0^{\be}x\ln\sin(\pi x) dx.
}

Letting $\be$ tend to unity in \betazeta,
\eqn\threezeta{
\ze'(0,\be)\rightarrow\ze'_{\rm rot}(0)-\ln(1-\be)-\ln2,
}
while from \difbetazeta
$$
\ze'(0,\be)\rightarrow 2\ze_R'(0)-\ln\pi-\ln(1-\be)+
2\int_0^1 x\ln\sin(\pi x) dx.
$$
The integral equals $-\ln\sqrt2$ and combining these two equations
reproduces the formula given above.
A similar discussion can be given for the general sphere.

There is virtually no extra work in obtaining the hemisphere results and
we place on record the first few expressions for $\ze_N'(0)-\ze_D'(0)$ from
which the requisite \zfs\ can be found by averaging with \resultsa,
$$\eqalign{
d=2,\quad\quad&2\ze_R'(0)\cr
d=3,\quad\quad&2\ze_R'(-1)+{1\over2}+\ln2\cr
d=4,\quad\quad&\ze_R'(-2)+2\ze_R'(0)+\ln3\cr
d=5,\quad\quad&{1\over3}\ze_R'(-3)+{11\over3}\ze_R'(-1)+{5\over18}+\ln4.\cr
}
$$
The two-hemisphere result is due to Weisberger [\Weisbergerb] and is
rederived in [\Dowka].
\vfill\eject
\sect{\bf 3. The conformal Laplacian.}

\noin A motivation for studying the conformal Laplacian is that, in the
case of the hemisphere, a conformal mapping from $\rS^d$ to $\oR^d$ enables
the Laplacian functional determinant in a $d$-ball to be found,
extending the $d=2$ result of Weisberger [\Weisbergerb]. (See also [\Dowkc].)
Unfortunately the relevant transformation has been worked out only for
$d=3$ and $d=4$, [\DandS]. The details of this specific application
will be given at another time.

For $\al=1/2$, the \zf\ cannot be written as a sum of the $H_p$ of \hdef, and
Vardi's Proposition 3.1 is not applicable. Our method proceeds
straightforwardly. In place of \neu\ and \dir\ we have
\eqn\neuc{
\ze_N'(0)=\ze_d'(0,d/2)+\ze_d'(0,d/2-1)-\sum_{r=1}^u{1\over2^{2r}r}
N^N_{2r}(d)\sum_{k=0}^{r-1}{1\over 2k+1}
}
\eqn\dirc{
\ze_D'(0)=\ze_d'(0,d/2)+\ze_d'(0,d/2+1)-\sum_{r=1}^u{1\over2^{2r}r}
N^D_{2r}(d)\sum_{k=0}^{r-1}{1\over 2k+1}.
}
Since $a$ remains the same, $N^N$ and $N^D$ are unaltered.

As in the previous section, it is convenient to discuss the sum, $\ze_{\rm
rot}=\ze_N+\ze_D$, and the difference, $\ze_{\rm diff}=\ze_N-\ze_D$,
separately. One has to treat odd and even $d$ individually.
Setting $d=2e$ or $d=2e+1$, we have for even $d$,
$$
2\ze_d(s,d/2)+\ze_d(s,d/2-1)+\ze_d(s,d/2+1)=
$$
\eqn\fourzetasco{\sum_{m=1}^\infty\left[2\comb{m+e-1}{2e-1}+\comb{m+e-2}
{2e-1}+\comb{m+e}{2e-1}\right]{1\over m^s}
}
while for odd $d$, this combination is
\eqn\fourzetasce{\sum_{m=0}^\infty\left[2\comb{m+e}{2e}+\comb{m+e-1}
{2e}+\comb{m+e+1}{2e}\right]{1\over (m+1/2)^s}
}

For the difference, $\ze_{\rm diff}$, the first terms in \fourzetasco\ and
\fourzetasce\ are removed and the signs of the last terms are reversed.

Rather than produce a general expression, the results for $\rS^3$,
$\rS^4$, $\rS^5$ and $\rS^6$ will be exhibited,
$$\eqalign{
d=3,\quad\quad&{5\over4}\ze_R'(-2)+\ze_R'(0)
-{1\over4}\ln2\approx-1.1303,\cr
\noalign{\vskip 5truept}
&{1\over8}-3\ze_R'(-1)+{1\over12}\ln2\approx0.6790.\cr
\noalign{\vskip 10truept}
d=4,\quad\quad&{2\over3}\ze_R'(-3)+{1\over3}\ze_R'(-1)
+{1\over144}\approx-0.04461,\cr
\noalign{\vskip 5truept}
&-\ze_R'(-2)\approx0.03045.\cr
\noalign{\vskip 10truept}
d=5,\quad\quad&{17\over96}\ze_R'(-4)+{5\over48}\ze_R'(-2)
-{1\over16}\ze'_R(0)+{1\over64}\ln2\approx0.06651,\cr
\noalign{\vskip 5truept}
&{1\over8}\ze_R'(-1)-{3\over8}\ze'(-3)-{5\over576}-
{11\over2880}\ln2\approx-0.03402.\cr
\noalign{\vskip 10truept}
d=6,\quad\quad&{1\over30}\big(\ze_R'(-5)-\ze_R'(-1)\big)
-{1\over1350}\approx0.00475,\cr
\noalign{\vskip 5truept}
&{1\over12}\big(\ze_R'(-2)-\ze'(-4)\big)\approx-0.00320.\cr
}
$$

\noin The first line gives $\ze'_{\rm rot}(0)$ and the second,
$\ze'_{\rm diff}(0)$. As $d$ becomes bigger, the values rapidly decrease
and oscillate about zero.

The $d=4$ wholesphere result agrees with that of Branson and \O rsted
[\Branson] who use Weisberger's method [\Weisbergera].

\sect{\bf 4. Lunes of any angle.}

\noin From now on, attention is confined to regions of the two-sphere.
Gromes, [\Gromes], gives the eigenvalues of $-\nabla^2$ for
Dirichlet conditions on a lune of angle $\be$ as $\la=\nu(\nu+1)$ with
\eqn\eigend{
\nu={\pi\over\be}m+n, \quad m=1,2\ldots,\quad n=0,1,\ldots\quad .
}
The eigenfunctions are $\sin(\mu\phi)\,P_\nu^{-\mu}(\cos\th)$ with
$\mu=m\pi/\be$. (See Pockels [\Pockels] pp100-109.)

For a half-lune the eigenvalues are as in \eigend\ except that $n$ ranges
over positive, odd integers. For Neumann conditions, $n$ is nonnegative
even. The discussion that follows uses Dirichlet conditions exclusively.

In terms of \genzetaa\ the lune \zf\ is
\eqn\lunezeta{
\ze_{\rm lune}(s)=\ze\big(s,1/2+\pi/\be,1/2;1,\pi/\be\big)
}
and that for the half-lune

\eqn\halflune{
\ze_{\rm halflune}(s) = \ze\big(s,3/2+\pi/\be,1/2;2,\pi/\be).
}

The analysis of [\Dowka] proceeds without change (but with $q$ replaced by
$\pi/\be$) up to the evaluation of the Barnes \zf\ and its derivative. For
arbitrary $\be$, a more involved continuation would be necessary but the
method of [\Dowka] can be extended to the case when $\be$ is a
{\it rational} part of $\pi$, $\be= q_1\pi/q_2$ \ie
$r_2=q_2/q_1$, with $q_1$ and $q_2$ integers. An interpolation of our previous
values can then be obtained.

A simple rescaling rewrites the \zf\ \genzetaa\ as
\eqn\genzetab{
\ze(s)\equiv\ze\big(s,a,\al;r_1,q_2/q_1\big)=
{r_1}^{-2s}\ze\big(s,a/r_1,\al/r_1;1,q_2/r_1q_1\big)
}
The ratio $q_2/r_1q_1$ is reduced to a relatively prime quotient, $e_2/
e_1$, and a further scaling gives
\eqn\genzetac{
\ze(s)\equiv\ze\big(s,a,\al;r_1,q_2/q_1\big)=
\left({e_1\over r_1}\right)^{2s}\ze\big(s,ae_1/r_1,\al e_1/r_1;e_1,e_2\big)
\equiv \left({e_1\over r_1}\right)^{2s}\ze_0(s).
}
$\ze_0$ can be treated by the methods of [\Dowka], the formulae from which
can be taken over directly, reinterpreting the quantities $a$, $\al$, $d_1$
and $d_2$ there as $e_1a/r_1$, $e_1\al/r_1$, $e_1$ and $e_2$.
For the lune, $r_1=1$ while for the half-lune, $r_1=2$; also $\al=1/2$.

{}From \genzetac\
\eqn\zetazero{
\ze(0)=\ze_0(0),\quad\ze'(0)=2\ln (e_1/r_1)\,\ze_0(0)+\ze'_0(0).
}

For the derivative in \zetazero\ we use the expression given in [\Dowka] in
terms of the Barnes \zf, $\ze_d$,
\eqn\zetazeroderiv{
\ze_0'(0)=\ze_2'(0,e_1+e_2\mid {\bf e})+
\ze_2'(0,e_1+e_2-e_1/r_1\mid{\bf e})-{\al^2e_1\over r_1^2e_2},
}
and for the derivative of the Barnes \zf\ the form involving a single
Hurwitz \zf,

\eqn\dzetaa{\eqalign{
\ze_2'(0,b\mid{\bf e})&={1\over
e_1e_2}\big(\ze_R'(-1,b)+\sum_i(\nu_i+b)\ln(\nu_i+b)\big)\cr
&+\sum_{\bf p}(1-w_b)\left(\ln\big(\Ga(w_b)/\sqrt(2\pi)\big)-(1/2-w_b)
\ln(e_1e_2)\right),\cr}
}
is preferred, despite the appearance of the (easily calculated) `missing
integers' $\nu_i$ described in [\Dowka].

In \dzetaa,
\eqn\wb{
w_b={b\over e_1e_2}+{p_1\over e_1}+{p_2\over e_2}
}
where the limits on the sums are $0\le p_1\le e_1-1$,
$0\le p_2\le e_2-1$.

The \zf\ at the origin needed in \zetazero\ is
\eqn\zetazerozero{
\ze_0(0)=\ze_2\big(0,e_2+e_1-e_1/2r_1\mid{\bf e}\big)+{\al^2e_1\over2r_1^2e_2}
}
and for $\ze_2(0)$
\eqn\barneszero{
\ze_2\big(0,b\mid{\bf e}\big)={1\over 12e_1e_2}\big(6b^2-6b(e_1+e_2)+
e_1^2+e_2^2+3e_1e_2\big).
}

Figure 2 shows the results for a half-lune. The effective action
$W_{\rm halflune}=-\ze'(0)/2$ is plotted against the angle $\be=\pi q_2/q_1$.
There is a shallow, local minimum just above $59\deg$
($\approx 59^\circ15'$).
\sect{\bf 4. Sectors of any angle.}

\noindent A stereographic projection of the half-lune $(2,2,\pi/\be)$ onto the
equatorial plane, from the north pole say, gives a sector of angle
$\be$ of the unit disc, if the $\be$-vertex is at the south pole.
Application of the L\"uscher-Symanzik-Weiss-Polyakov conformal
transformation, [\Luscher, \Polyakov, \Bukhb, \Alvarez], enables one to
obtain the functional determinant on such a sector
from that on the half-lune. Rescaling the sector and projecting back onto the
sphere allows the functional determinant on a more general $(2,2,\pi/\be)$
spherical triangle to be determined. We pursue this program in this section.

The results of [\Dowkc] allow the effective action of a sector of angle
$\be$ to be written

\eqn\usectora{
W_{\rm sector}=W_{\rm halflune}-{\be\over6\pi}+{1\over24}\left({\pi\over\be}+
{\be\over\pi}\right)\ln(2/a)-{1\over8}\ln a.
}
In order to change the radius to $a$, an amount $\ln a\,\ze(0)$ has been
subtracted.
$\ze(0)$ is given by \zetazerozero\ with $r_1=2$,
\eqn\hlzetazero{
\ze(0)=\ze_0(0)={1\over8}+{1\over24}\left({\pi\over\be}+{\be\over\pi}\right).
}
Figure 3 shows a plot of $W_{\rm sector}$, for $a=1$, against angle. Of course,
in the computation, the angle $\be$ has been taken to be a rational part of
$\pi$.

Having found the effective action on a disc-sector of arbitrary radius, we
place its centre at the origin and stereographically project back
onto the unit sphere thereby obtaining an isosceles spherical triangle,
$(2,2,\pi/\be)$ with angle $\be$ at the south pole.

A slight modification of the calculation in [\Dowkc] for a cap, allowing for
the different range of angle integration, yields
$$
W_{\rm triangle}(\be,\si)=W_{\rm sector}(\be,\si)-
{\be\over24\pi}\big(4\ln2-\si+\si(1-\si)\ln(2-\si)+3\si^2-9\si+3\big)$$
\eqn\trianglea{
-{1\over24}\big({\pi\over\be}-{\be\over\pi}\big)\ln2 +{1\over8}\ln(2-\si)
}
where the final two terms come from the corners.
The area of the triangle is $\be\si$ and it has sides $\ga$, $\ga$ and
$\be\sin\ga$ where $\si=1-\cos\ga$. Also $\si=a^2/(1+a^2)$ and $\ka=\cot\ga$.

Combined with \usectora, \trianglea\ becomes
$$
W_{\rm triangle}(\be,\si)=W_{\rm halflune}(\be)+{1\over48}\big(2\be\si(2-\si)
-3+{\pi\over\be}-{\be\over\pi}\big)\ln(2-\si)-$$
\eqn\triangleb{
{1\over48}\big({\pi\over\be}-{\be\over\pi}+4\big)\ln\si+{\be\over24\pi}
(3\si-7)(\si-1).
}

This expression always diverges as $\si$ tends to $0$ and, except at the angle
$\be\approx 54^\circ30'$, also as $\si\rightarrow2$.
Figure 4 shows a contour plot of the effective action as a function of the
angle $\be$ and the variable $\si$. There is a saddle point at
$\be\approx90^\circ$, $\si\approx1.64$.

Perhaps it should be remarked that in the $\si=2$ limit, the triangle does not
become a lune of angle $\pi/\be$, except in a crude, pictorial sense. The
$\be$ vertex
is not approximated by two right angles joined by an infinitesimal arc of
latitude, which is what the third side of the triangle tends to as it
approaches the north pole.

\sect{\bf 5. Comments.}

\noin
The discussion of the wholesphere, and the hemisphere, shows, we think, that
our method, [\Dowka], has some technical and interpretative advantages over
the others available, [\Hortacsu,\Weisbergera,\Vardi]. Our technique is closest
in spirit to that of Weisberger [\Weisbergera].

A continuation of the lune \zf\ for arbitrary $\be$ is
called for and can be derived using the standard technique of Euler summation.
This adds nothing numerically to what we already have but is intellectually
more satisfying.
\vfill\eject
\sect{FIGURE CAPTIONS.}
\vskip 15truept
\noin Figure 1. Derivative at zero, $F$, of the \zf\ on a sphere plotted
against sphere dimension, $d$. The dotted line joins even-$d$ values.
\vskip 15truept
\noin Figure 2. Effective action for the half-lune $(2,2,\pi/\be)$ plotted
against angle $\be=q_2\pi/q_1$ with $q_1,\,q_2\in\oZ$.
\vskip 15truept
\noin Figure 3. Effective action for a sector of angle $\be$ of the unit disc.
\vskip 15truept
\noin Figure 4. Effective action for an isosceles spherical triangle against
$\si$ and included angle $\be$.
\vskip 15truept
\vfill\eject

  \immediate\closeout\reffile
  \noindent{{\bf References}}\bigskip\frenchspacing

  \input refs.tmp\vfill\eject\nonfrenchspacing
\end